\documentclass[twocolumn,showpacs,prb]{revtex4}
\usepackage{graphicx}

\begin{document}
\title{Real Space Hartree-Fock Configuration Interaction Method For 
Complex Lateral Quantum Dot Molecules}

\author{Ramin M. Abolfath}

\author{Pawel Hawrylak}

\affiliation{
Institute for Microstructural Sciences,
National Research Council of Canada,
Ottawa, K1A 0R6, Canada,
}

\date{\today}

\begin{abstract}
We present  unrestricted Hartree Fock method 
coupled with configuration interaction (CI) method 
(URHF-CI) suitable for the calculation
of ground and excited states of large number of electrons localized by
complex gate potentials in quasi-two-dimensional quantum dot molecules. 
The method employs real space finite difference method,
incorporating strong magnetic field, 
for the calculating of single particle states. The  Hartree-Fock 
method is employed for the calculation of direct and exchange
interaction contribution to the ground state energy. The effects of
correlations are included in energies and directly in the many-particle wavefunctions
via configuration interaction (CI) method using a limited set of 
excitations above the Fermi level.  
The   URHF-CI method and its performance are  illustrated 
on the example of  ten electrons confined in a two-dimensional
quantum dot molecule.
\end{abstract}
\maketitle

\section{Introduction}

There is currently interest in localizing electrons in quantum dots 
and utilizing their spin for quantum information processing
\cite{qinfo,brum,Loss-DiVincenzo}.
Alternatively, we can view such devices as highly tunable quantum dot molecules,
with large numbers of electrons or qubits, with gates and magnetic field
used to perform quantum operations. Even with a relatively small number of bits,
to understand working of such devices one needs
a computational tool suitable for the calculation of wavefunctions 
of at least tens of electrons confined by complex gate potentials
and in magnetic field. The wavefunctions must capture entanglement, 
or in condensed matter language, correlations. The configuration 
interaction (CI) method is a suitable candidate  modified
here for the problem of quantum dot molecules.
In the configuration interaction (CI) method the Hamiltonian of an
interacting system is calculated in the basis of a finite number of 
many-electron configurations and diagonalized exactly. 
Exact diagonalization (ED) has been an important tool 
in the context of quantum chemistry of atoms and molecules \cite{Szabo_book}.
In condensed matter physics ED has been used to
investigate the electronic and optical properties of 
artificial atoms and molecules in quantum
dots (QDs) \cite{jacak98, maksym90, merkt91, pfannkuche93, hawrylak93,
yang93, hawrylak93b, palacios94, wojs95, oaknin96, wojs96, maksym96,
hawrylak96, wojs97, wojs97b, eto97, eto97b, maksym98, imamura98, imamura98b,
hawrylak99, creffield99, bruce00, reimann00, mikhailov02,
palacios,KRKM,HuDasSarma,WKH,WKH2,Rontani,RaminWojtekPawel}
and finite size quantum Hall systems.
\cite{Yoshioka,HaldaneRezayi,RezayiHaldane,FanoOrtolaniColombo}
The direct access to electron correlations in 
the many-particle wave functions calculated by CI 
makes CI useful for the purpose of quantitative
description of the electron entanglement in the
context of solid state quantum information architectures.
The drawback of the method is its slow convergence and applicability to
a small number of electrons. While  mean-field theories such as
Hartree-Fock (HF) \cite{yannouleas99,reusch01},
and density-functional theory (DFT) 
\cite{koskinen97, austing99, hirose99,
steffens99, wensauer00, wensauer01,Zhuang}
can be applied to large electron numbers, they neglect electron correlations,
or entanglement, in the wavefunction.

To include electron correlations explicitly in the many-particle 
wavefunctions one may follow quantum chemistry methods which 
start with HF or DFT calculations of effective single particle orbitals, and
use them in the construction of a finite number of many-electron configurations
as input in configuration interaction 
calculations \cite{Szabo_book}.  These ideas have been already explored for the calculation
of electronic properties of quantum dots with large electron numbers\cite{WKH}
and quantum dot molecules.\cite{HuDasSarma,RaminWojtekPawel}

The objective of this work is to present an efficient HF-CI method: (a) based
on real space calculations of single particle states and hence suitable for
complex quantum dot potentials, (b) combining complex potentials with 
strong magnetic field, and
(c) capable of treating correlations (entanglement) in electron wavefunction 
for large number of
electrons.
The   URHF-CI method and its performance are  illustrated 
on the example of  ten electron confined in a two-dimensional
quantum dot molecule.

The paper is organized as follows:
In section \ref{SecII} we  present the Hamiltonian of 
electrons confined in the lateral gated quantum dot molecule.
The calculation of the single particle spectrum,  including
the details of the implementation of the localized gauge, is
presented in section \ref{SecIII1}.
In section \ref{SecIV} our many electron approach, including the
unrestricted Hartree-Fock approximation, is introduced.
The central results of this study, the calculation of the ground
state energy and excitation gap by configuration interaction method
using unrestricted Hartree-Fock
 URHF-CI basis and non-interacting single particle SP-CI basis, are presented in
section \ref{SecV}. The paper is summarized in section \ref{SecVI}.

\section{Hamiltonian}
\label{SecII}

We consider $N$ electrons  with effective mass $m^*$, moving on a plane $(x,y)$
defined by the GaAs/GaAlAs interface, with complicated  confining potential
$V(\vec{r})$ as a function of electron position $\vec{r}$,
created by metallic gates on the surface of the heterostructure.
The magnetic field, defined by vector potential $A$, is perpendicular to the 2D plane.
The effective $N$-electron quantum dot molecule Hamiltonian in the presence of
an external magnetic field can be written as:
\begin{eqnarray}
H &=& \sum_{i=1}^N \left[ \frac{1}{2m^*} 
\left(\frac{\hbar}{i}\vec{\nabla}_i + \frac{e}{c} A(\vec{r}_i)\right)^2 
+ V(x_i,y_i) \right] \nonumber \\&&
+ \frac{e^2}{2\epsilon}\sum_{i \neq j}\frac{1}{|\vec{r}_i - \vec{r}_j|},
\label{MpH}
\end{eqnarray}
where $e$ is electron charge, $c$ is velocity of light, and
$\epsilon$ is the host semiconductor dielectric constant.The Zeeman splitting (very small for GaAs) is neglected here.
The first term in Eq. (\ref{MpH}) is the sum of single particle Hamiltonians and the second term is
the Coulomb repulsion. The electron-electron repulsion is screened by the dielectric constant,
which for GaAs is $\epsilon\approx 12.4$. 

In what follows, we present the results in effective atomic units with  
length unit in effective Bohr radii $a^*_0 = (\hbar^2/2m^*)/(e^2/2\epsilon)$, 
and energy unit in effective Rydberg $Ry^* = e^2/(2\epsilon a^*_0)$.
Using material parameters for GaAs with $m^* \approx 0.067 m_0$ 
($m_0$ is the bare electron  mass)
$Ry^*=5.93 meV$, and $a^*_0 = 9.79 nm$.

\section{Single particle spectrum in a magnetic field}
\label{SecIII1}

\begin{figure}
\begin{center}\vspace{1cm}
\includegraphics[width=0.7\linewidth]{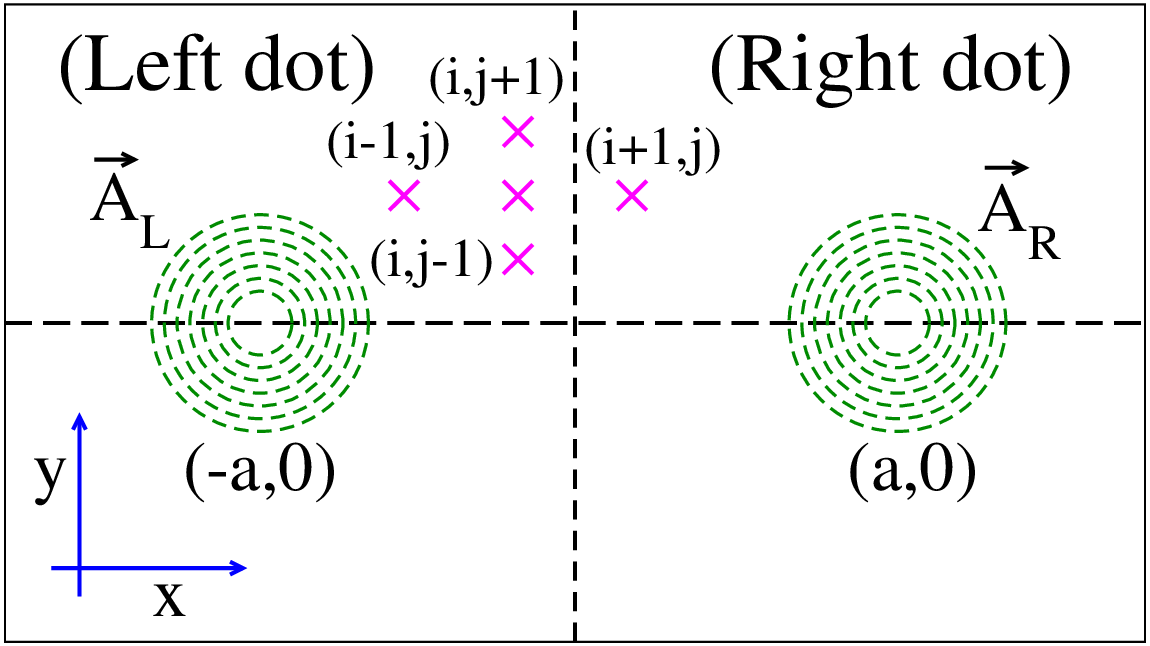}
\caption{discrete points in left (L) and right (R) dot regions.}
\label{comp_box}
\end{center}
\end{figure}

The single particle Schr\"odinger equation is given by
\begin{eqnarray}
\left[ \frac{1}{2m^*} \left(\frac{\hbar}{i}\vec{\nabla} + \frac{e}{c} A(\vec{r})
\right)^2 + V(\vec{r})\right] \tilde{\varphi}(\vec{r}) 
= \varepsilon \tilde{\varphi}(\vec{r}),
\label{LSE}
\end{eqnarray}
where $A=(B/2)(-y,x)$ is the vector potential seen by electrons, $V(\vec{r})$
is the confining potential, and $B$ is the magnetic field.
For a single quantum dot the potential is often well
approximated by an analytical form $V(\vec{r})=1/2 m \omega^2 r^2$.
For quantum dot molecules  the potential will have several minima in 
which to localize electrons, and the most convenient way to solve the single particle 
problem is by discretizing the 
Schr\"odinger equation. 
As an example of a complicated molecular quantum dot potential we study
a double Gaussian potential
$V(x,y)=V_L~ \exp[{-\frac{(x+a)^2+y^2}{\Delta^2}}]
        +V_R~ \exp[{-\frac{(x-a)^2+y^2}{\Delta^2}}]
+V_p \exp[{-\frac{x^2}{\Delta_{Px}^2}-\frac{y^2}{\Delta_{Py}^2}}]$,
with $V_L$ and $V_R$ measuring the depth of the left and right dot, and 
$V_p$ controlling the height of the tunneling barrier. The centers of the 
two dots are separated by $2a$. 
To describe the typical coupled quantum dots we use the parameters
$V_L=V_R=-10, a=2, \Delta=2.5$, and $\Delta_{Px}=0.3$, 
$\Delta_{Py}=2.5$, in effective atomic units.
$V_p$ which controls the central potential barrier mimics the plunger 
gate strength and is varied between zero and $10 Ry^*$, independent 
of the locations of the quantum dots.

The choice of gauge $A$ plays significant role in
improving the numerical accuracy of single particle spectrum. 
We  use 
a gauge field which is adopted to the geometry of the
confining potential. For a double dot molecule with two minima,
 we divide the plane of coupled
quantum dots into left and right domains, as shown in Fig. \ref{comp_box}.
In order to end up with a well defined left and right quantum dots 
with increasing barrier height, i.e. with dots with gauge centered in the origin,
we define left (right) wavefunctions and Hamiltonians,
and carry out gauge transformation on both Hamiltonians and wavefunctions. 
The resulting vector  potentials localized in centers of each
dot are $A_L=(B/2)(-y,x+a)$ for the left dot ( $x<0$ ),
 and  $A_R=(B/2)(-y,x-a)$ for the right dot ($x>0$).
 Introducing corresponding wavefunctions in the left (right) dot 
 $\tilde{\varphi}_{i,j}= \tilde{\varphi}^L_{i,j} \exp(+ i \omega_c ay/4)$
 ($\tilde{\varphi}_{i,j}=\tilde{\varphi}^R_{i,j} \exp(- i \omega_c ay/4)$)
allows us to write an explicit form of the
 discrete form of Schr\"odinger equation. 
In the left-dot the Schr\"odinger equation
can be obtained by expanding $\tilde{\varphi}^L$ around the point $(i,j)$ 
\begin{eqnarray}
&&
-\frac{1}{h^2} 
\left[\tilde{\varphi}^L_{i+1,j} + \tilde{\varphi}^L_{i-1,j} 
+ \tilde{\varphi}^L_{i,j+1} + \tilde{\varphi}^L_{i,j-1} 
- 4 \tilde{\varphi}^L_{i,j}\right] \nonumber \\&&
-i\frac{\hbar\omega_c}{4h}
\left[(x_i + a) (\tilde{\varphi}^L_{i,j+1} - \tilde{\varphi}^L_{i,j-1}) 
     - y_j (\tilde{\varphi}^L_{i+1,j} - \tilde{\varphi}^L_{i-1,j})\right] 
     \nonumber \\&&
+ \frac{(\hbar\omega_c)^2}{16}\left[(x_i+a)^2+y_j^2\right]\tilde{\varphi}^L_{i,j}
+ V_{i,j} \tilde{\varphi}^L_{i,j} = \varepsilon \tilde{\varphi}^L_{i,j} 
\label{dSE}
\end{eqnarray}
where $x_i = i h$, and $y_j = j h$, and $h$ is the grid spacing.
A similar equation holds for the wavefunction in the right dot. 

As it is illustrated in Fig. \ref{comp_box},
adjacent to the inter-dot boundary,
the point $(i,j)$ in L, is connected to 
the point $(i+1,j)$ in R, via Schr\"odinger equation.
In this case $\tilde{\varphi}^L_{i+1,j}$ in Eq.(\ref{dSE}) is already in the right dot and
is not known and must be replaced by
$\tilde{\varphi}^R_{i+1,j} \exp(-i \omega_c ay/2)$.

One should note that this separation  of gauge fields into left and right  leads to
a uniform magnetic field $B=\nabla\times A$ in both  half-planes,
but it produces an artifical  discontinuity in $B$ along $x=0$.
To avoid this unphysical discontinuity, 
in finite difference method, the grids along the 
$x$-axis have been set slightly away from $x=0$ such that
$x=0$-line has been excluded from the real space 
finite difference calculation.

\begin{figure}
\begin{center}\vspace{1cm}
\includegraphics[width=0.9\linewidth]{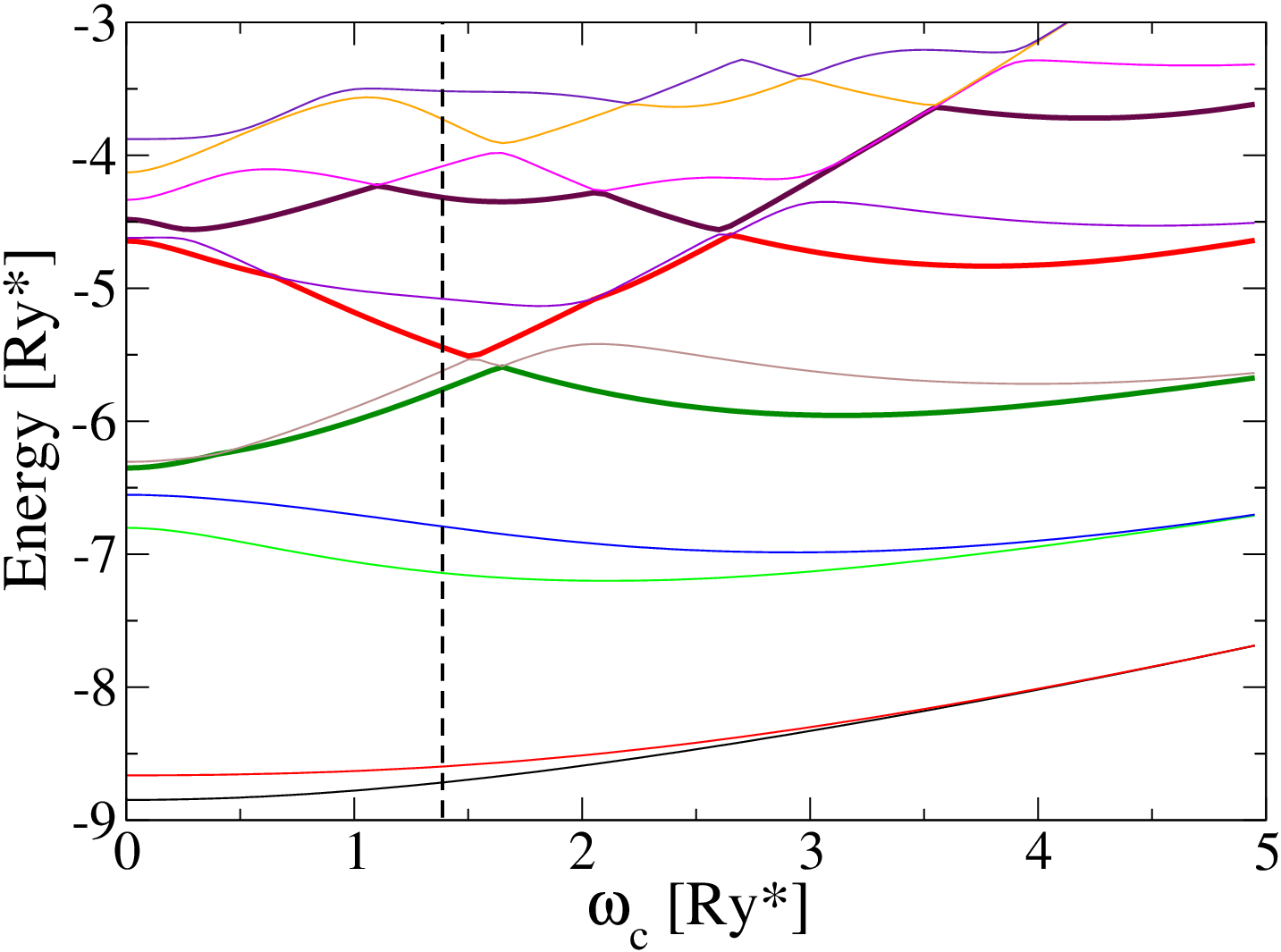}
\caption{Double dot single particle spectrum vs. cyclotron energy.
Several crossing between Landau levels is visible at 
$\hbar\omega_c=1.4Ry^*$.
}
\label{E_sp}
\end{center}
\end{figure}

The single particle spectrum $\tilde{\epsilon}_j$ and $\tilde{\varphi}_j$
is calculated accurately on a finite mesh of $N_x \times N_y$ 
($102\times 51$ mesh points
are used in what follows) using conjugated gradient methods.
The magnetic field dependence of single particle spectrum
calculated by this method, is shown in Fig. \ref{E_sp}.
At zero magnetic field Fig. \ref{E_sp} shows the formation of 
hybridized S, P, and D shells.
In high magnetic field we observe the formation of
shells of closely spaced pairs of levels (with opposite parity). 
In intermediate magnetic field states with opposite parity cross
and states with the same parity anticross. The crossing of many levels
at e.g. 
$\hbar\omega_c=1.4Ry^*$, marked with dashed line in Fig.\ref{E_sp},
leads to the most  computationally challenging many-electron problem.

\section{Many electron spectrum}
\label{SecIV}

The single-particle (SP) states calculated in previous section can be used as a basis in
configuration-interaction (CI) calculation.
Denoting the creation (annihilation) operators for electron in
non-interacting SP state $|\alpha\sigma\rangle$ by 
$\tilde{c}^\dagger_{\alpha\sigma}~(\tilde{c}_{\alpha\sigma})$,
the Hamiltonian of an interacting system in second quantization
can be written as
\begin{eqnarray}
H&=&\sum_{\alpha}\sum_{\sigma} 
\tilde{\epsilon}_{\alpha} 
\tilde{c}^\dagger_{\alpha\sigma} \tilde{c}_{\alpha\sigma} 
\nonumber \\ &&
+ \frac{1}{2}\sum_{\alpha\beta\gamma\mu}\sum_{\sigma\sigma'}
\tilde{V}_{\alpha\sigma,\beta\sigma',\gamma\sigma',\mu\sigma}
\tilde{c}^\dagger_{\alpha\sigma} \tilde{c}^\dagger_{\beta\sigma'} 
\tilde{c}_{\gamma\sigma'} \tilde{c}_{\mu\sigma} 
\label{multiparticle}
\end{eqnarray}       
where the first term is the single particle Hamiltonian,
and
$\tilde{V}_{\alpha\sigma,\beta\sigma',\mu\sigma',\nu\sigma} = 
\int d\vec{r} \int d\vec{r'}  
\tilde{\varphi}^*_{\alpha\sigma}(\vec{r})
\tilde{\varphi}^*_{\beta\sigma'}(\vec{r'}) 
\frac{e^2}{\epsilon|\vec{r}-\vec{r'}|}
\tilde{\varphi}_{\mu\sigma'}(\vec{r'})
\tilde{\varphi}_{\nu\sigma}(\vec{r})$, 
is the two-body Coulomb matrix element.

Alternatively we may use the single particle states to first construct the
Hartree-Fock orbitals.
In this scheme electrons are treated as independent particles moving 
in a self-consistent HF field.
Similar to single particle configuration-interaction (SP-CI) method,
the HF orbitals can be used as a basis of the interacting Hamiltonian 
in second quantization, and subsequently in the CI calculation.
In this method electron-electron interactions are included
in two steps:
direct and exchange interaction using Hartree-Fock approximation,
and correlations using HF basis in the configuration interaction method.

\subsection{Unrestricted Hartree-Fock Approximation (URHFA)}

Hartree-Fock approximation is  a mean field approach to many
body systems which accounts for the 
direct and exchange Coulomb interactions.
Combining HFA with more sophisticated many body methods (such as CI)
allows to isolate the effect introduced by correlations.   
The Hartree-Fock ground state (GS) of the electrons 
with given $S_z = (N_\uparrow - N_\downarrow)/2$
is a single Slater determinant

\begin{equation}
|\Psi\rangle = 
\frac{1}{\sqrt{N}!}\sum_{P}(-1)^P
\varphi_{1\sigma_1}(r_{P1})|\sigma_1\rangle \dots
\varphi_{N\sigma_N}(r_{PN})|\sigma_N\rangle,
\end{equation}  
with $P$ the 
permutation operator.
The HF orbitals $|\varphi_{i\sigma}\rangle$ which describe the state of a 
dressed quasi-particle in the quantum dot molecule can be determined by 
minimizing the HF energy $E_{HF} = \langle \Psi|H|\Psi\rangle$ with respect to 
$\varphi_{i\sigma}(r)$
\begin{eqnarray}
&&\left\{\hat{T}+\int d^2r'\sum_{\sigma'}\sum_{j=1}^{N_{\sigma'}}
|\varphi_{j\sigma'}(\vec{r}')|^2
V(|\vec{r}-\vec{r}'|) \right\}
\varphi_{i\sigma}(\vec{r}) \nonumber \\ &&
- \sum_{\sigma'}\sum_{j=1}^{N_{\sigma'}}\varphi_{j\sigma}(\vec{r})\int d^2r'
\varphi^*_{j\sigma'}(\vec{r}') V(|\vec{r}-\vec{r}'|)
\varphi_{i\sigma'}(\vec{r}')\delta_{\sigma\sigma'}\nonumber \\ &&
=\epsilon_{i\sigma}\varphi_{i\sigma}(\vec{r}).
\label{HFeq}
\end{eqnarray}
$\hat{T}=\frac{1}{2m^*}(\frac{\hbar}{i}\vec{\nabla} + \frac{e}{c} A)^2$ 
is the non-interacting single particle Hamiltonian, 
and $V(r)=e^2/\epsilon r$ is the  Coulomb interaction.
This equation is  a set of coupled equations,
for spin up and spin down states.
To find numerical solutions of the Hartree-Fock equations
we expand the HF orbitals $|\varphi_{i\sigma}\rangle$
in terms of single particle states $|\tilde{\varphi}_{\alpha}\rangle$:
$|\varphi_{i\sigma},\sigma\rangle = \sum_{\alpha=1}^{N_l} a^{(i)}_{\alpha\sigma} 
|\tilde{\varphi}_{\alpha},\sigma\rangle$.
This transforms the HF equation, Eq.(\ref{HFeq}), to 
the self-consistent Pople-Nesbet equations \cite{Szabo_book}:
\begin{eqnarray}
&&\sum_{\gamma=1}^{N_l}
\{\tilde{\epsilon}_\mu\delta_{\gamma\mu}+
\sum_{\alpha,\beta=1}^{N_l}\tilde{V}_{\mu\alpha\beta\gamma}
\sum_{\sigma'}
\sum_{j=1}^{N_{\sigma'}}
a^{*(j)}_{\alpha\sigma'} a_{\beta\sigma'}^{(j)} 
\nonumber \\&&
-\tilde{V}_{\mu\alpha\gamma\beta}
\sum_{\sigma'}\sum_{j=1}^{N_{\sigma'}}
a^{*(j)}_{\alpha\sigma'} a_{\beta\sigma'}^{(j)}\delta_{\sigma,\sigma'}  \} 
a_{\gamma\sigma}^{(i)} 
= \epsilon_{i\sigma} ~ a_{\mu\sigma}^{(i)},
\label{urhfeq1}
\end{eqnarray}
where $\tilde{V}_{\alpha\beta\mu\nu}$ are Coulomb matrix elements
calculated using non-interacting single particle states.

This procedure results in $N_l$ HF states.
The $N$-lowest energy states form a Slater determinant
occupied by HF (quasi) electrons corresponding to HF ground state.
The rest of orbitals with higher energies are 
outside of the HF Slater determinant (unoccupied states).

\begin{figure}
\begin{center}
\includegraphics[width=0.9\linewidth]{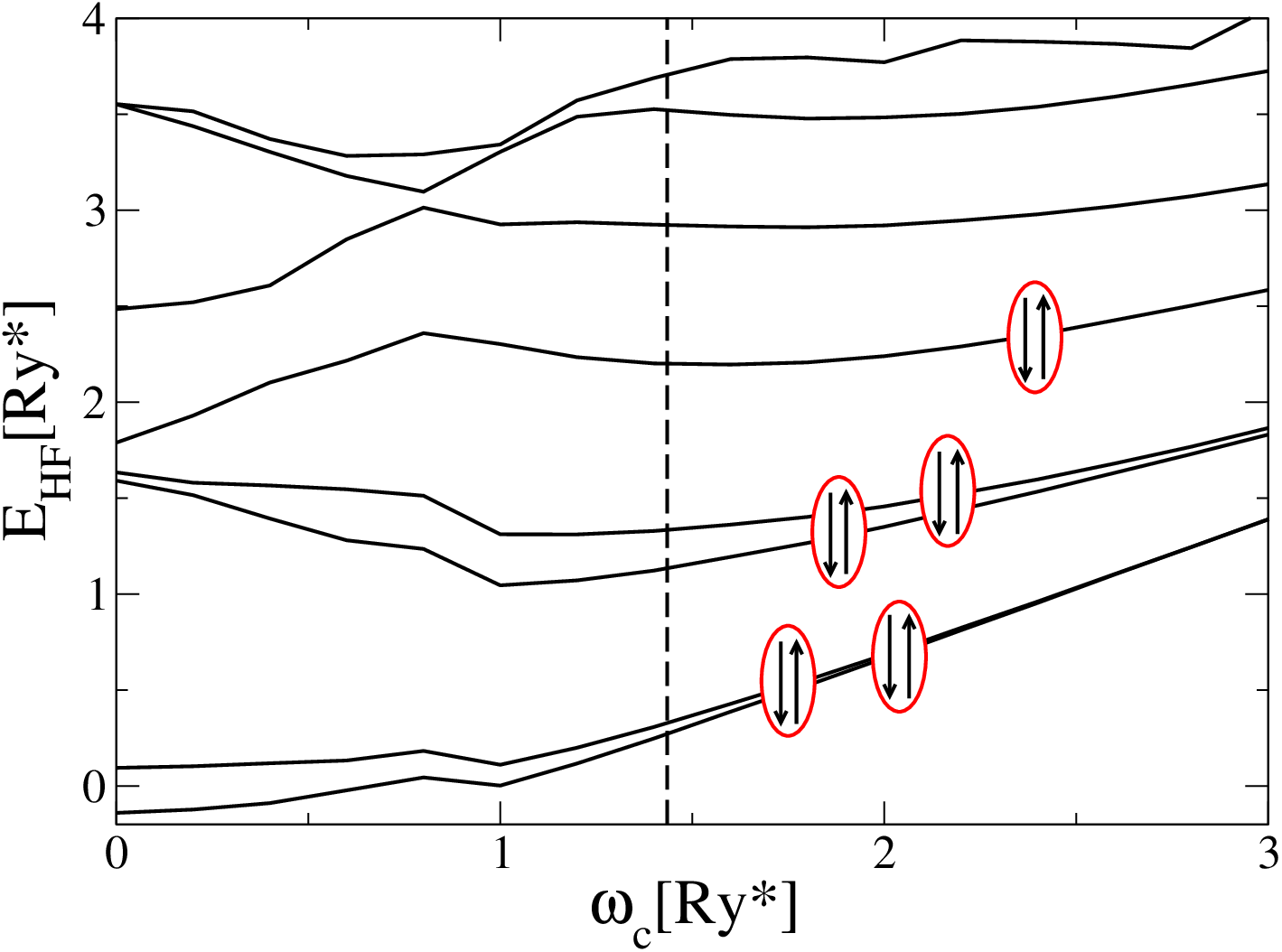}\vspace{1cm}
\noindent
\caption{
URHF eigen-energies vs. cyclotron energy for 10 electrons
with $S_z=0$.
The HF energy gap between HOMO (highest occupied molecular orbital),
and LUMO (lowest unoccupied molecular orbital) is clearly visible.
A comparison between SP-CI and URHF-CI calculation will be presented
at $\hbar\omega_c=1.4Ry^*$, and it is shown by the dash line.
}
\label{E_HFvsB}
\end{center}\vspace{0.5cm}
\end{figure}

The calculated HF eigen-energies  
for the $N=10$ electrons with $S_z=0$  as a function of 
magnetic field are shown in Fig. \ref{E_HFvsB}.
Comparing HF spectrum (Fig. \ref{E_HFvsB}) with the single particle 
spectrum (Fig. \ref{E_sp}),
one observes that a HF gap developed at the Fermi level,
between the highest occupied molecular state, and the lowest
unoccupied molecular state, and the Landau level crossing 
between single particle Landau levels has been shifted to lower 
magnetic fields.

\begin{figure}
\begin{center}
\includegraphics[width=0.9\linewidth]{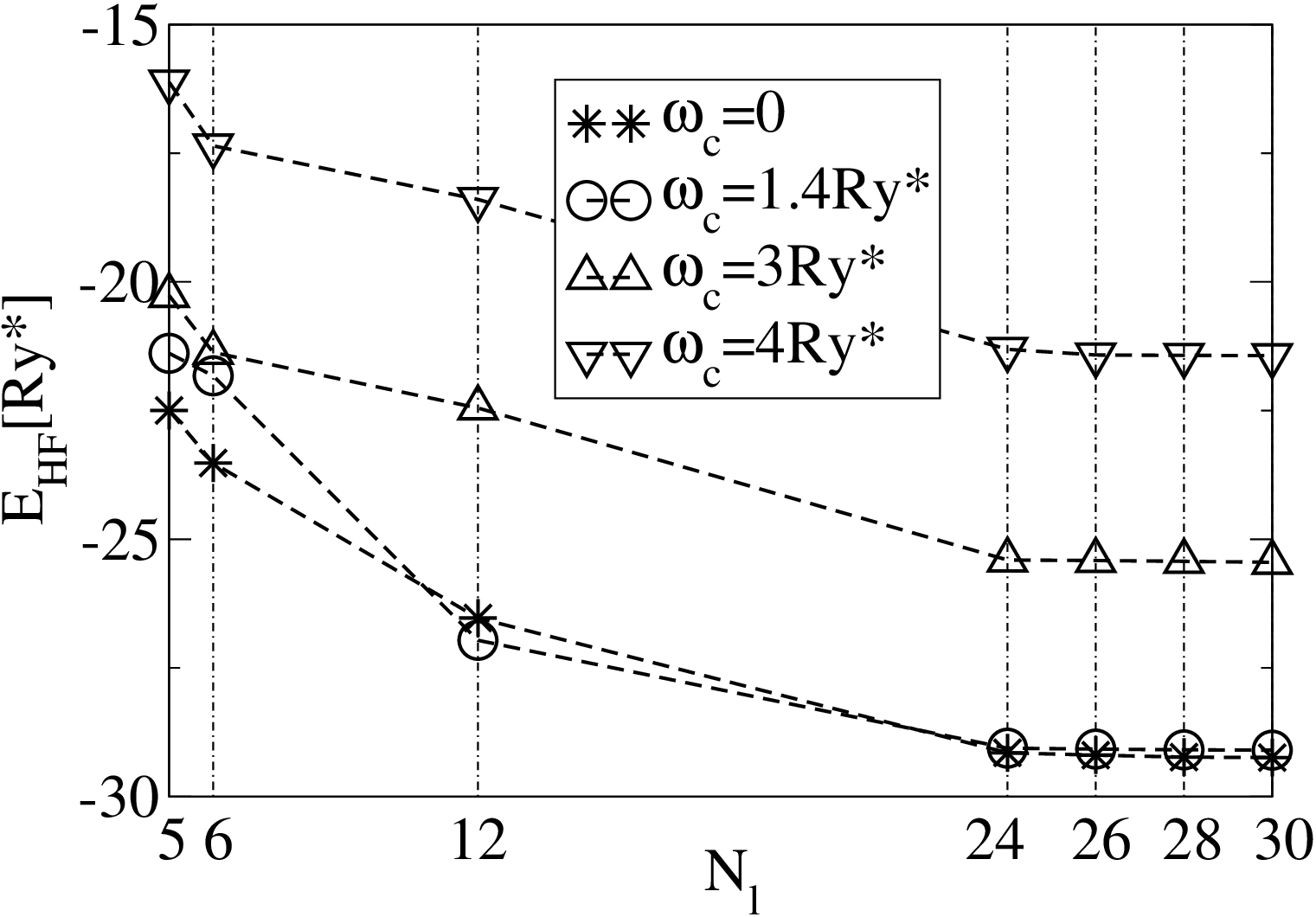}
\noindent
\caption{
HFA ground state energy 
vs. number of single particle states participated in HF calculation
($N_l$) for 10 electrons.
}
\label{EHFvsN}
\end{center}
\end{figure}

Fig. \ref{EHFvsN} illustrates the convergence of  HF
energy with respect to the number $N_l$ of single particle orbitals
used in HF expansion.

In the following we study the effect of correlations by
including quasielectron-quasihole excitations in the HF wave functions 
using CI method.

\section{Configuration-interaction method}
\label{SecV}

\begin{figure}
\begin{center}\vspace{1cm}
\includegraphics[width=0.9\linewidth]{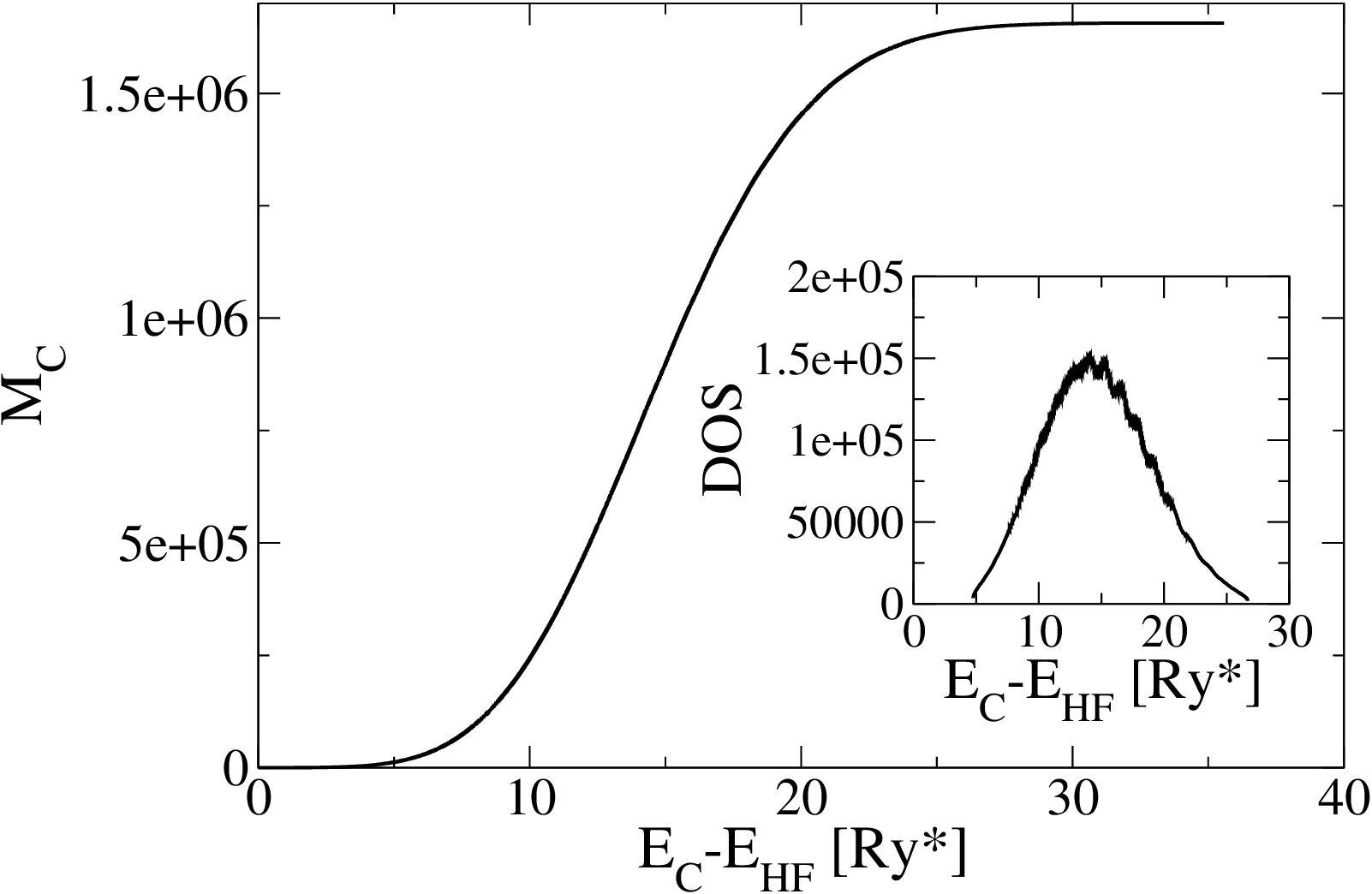}\vspace{1cm}
\noindent
\caption{URHF electron-hole excitation
energies $E_C = \langle\Psi_{M_C}|H|\Psi_{M_C}\rangle$ 
for $N_s=13$ corresponding to $N_C=1656369$ is plotted
vs. $M_C$.
Here $1\leq M_C\leq N_C$ is an integer number. 
$|\Psi_{M_C}\rangle$ is the Slater determinant of the configuration
$M_C$th, formed by HF basis.
$E_C=E_{HF}$ (the HF ground state) if $M_C=1$.
The URHF density of states ($dM_C/dE_C$) is shown in inset.
Its peak is below the HF energy mean value.
}
\label{DOS8}
\end{center}
\end{figure}

\begin{figure}
\begin{center}\vspace{1cm}
\includegraphics[width=0.9\linewidth]{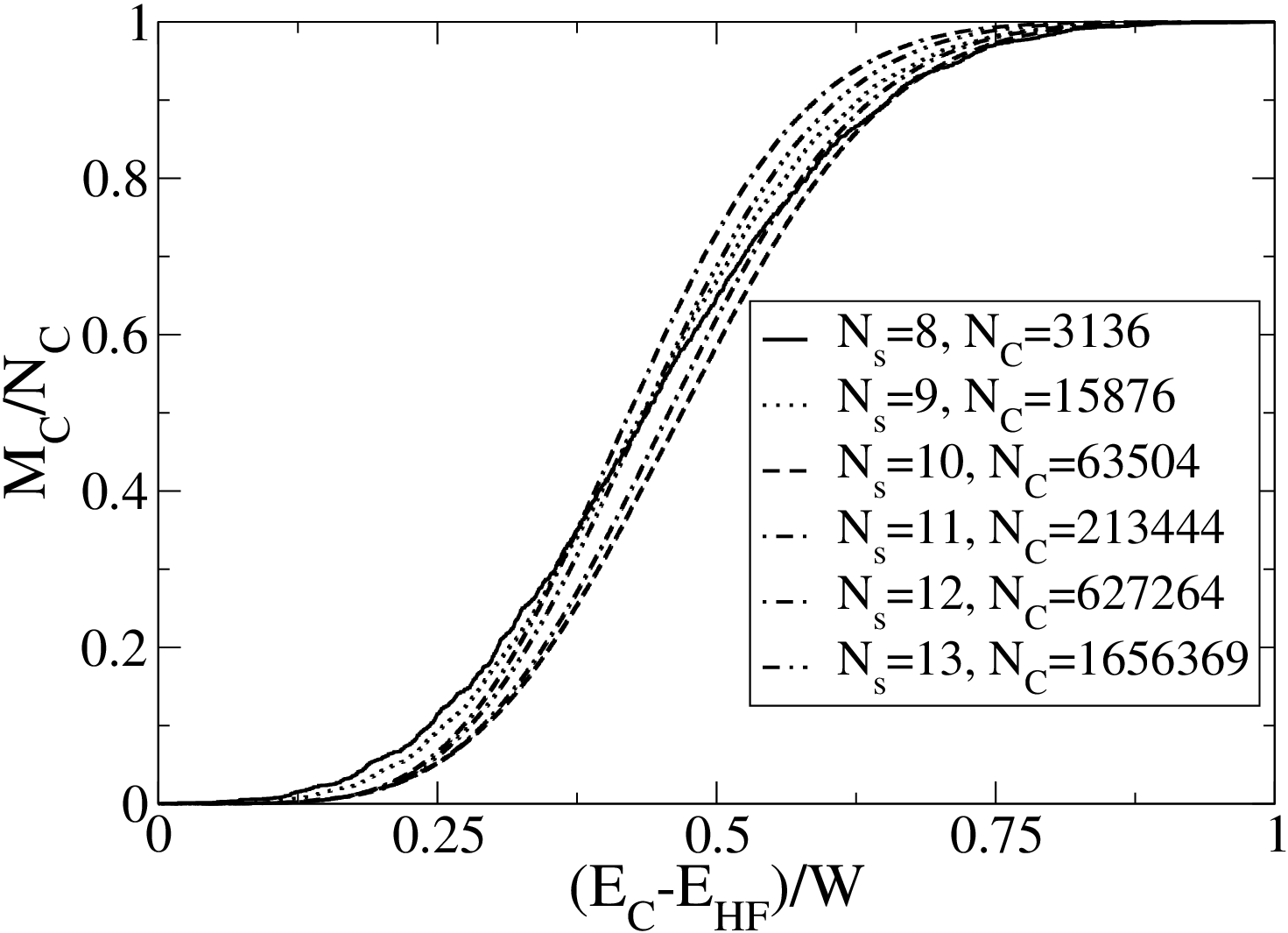}
\noindent
\caption{The URHF excitation energies $E_C$ corresponding to the
$M_C$th diagonal matrix elements in $H$ (see Eq.\ref{IR7}), 
are shown for various configurations. 
Due to the finite band width $W$ of the HF spectrum,
they approximately follow a universal functional form.
}
\label{Scaling}
\end{center}
\end{figure}

\begin{figure}
\begin{center}\vspace{1cm}
\includegraphics[width=0.9\linewidth]{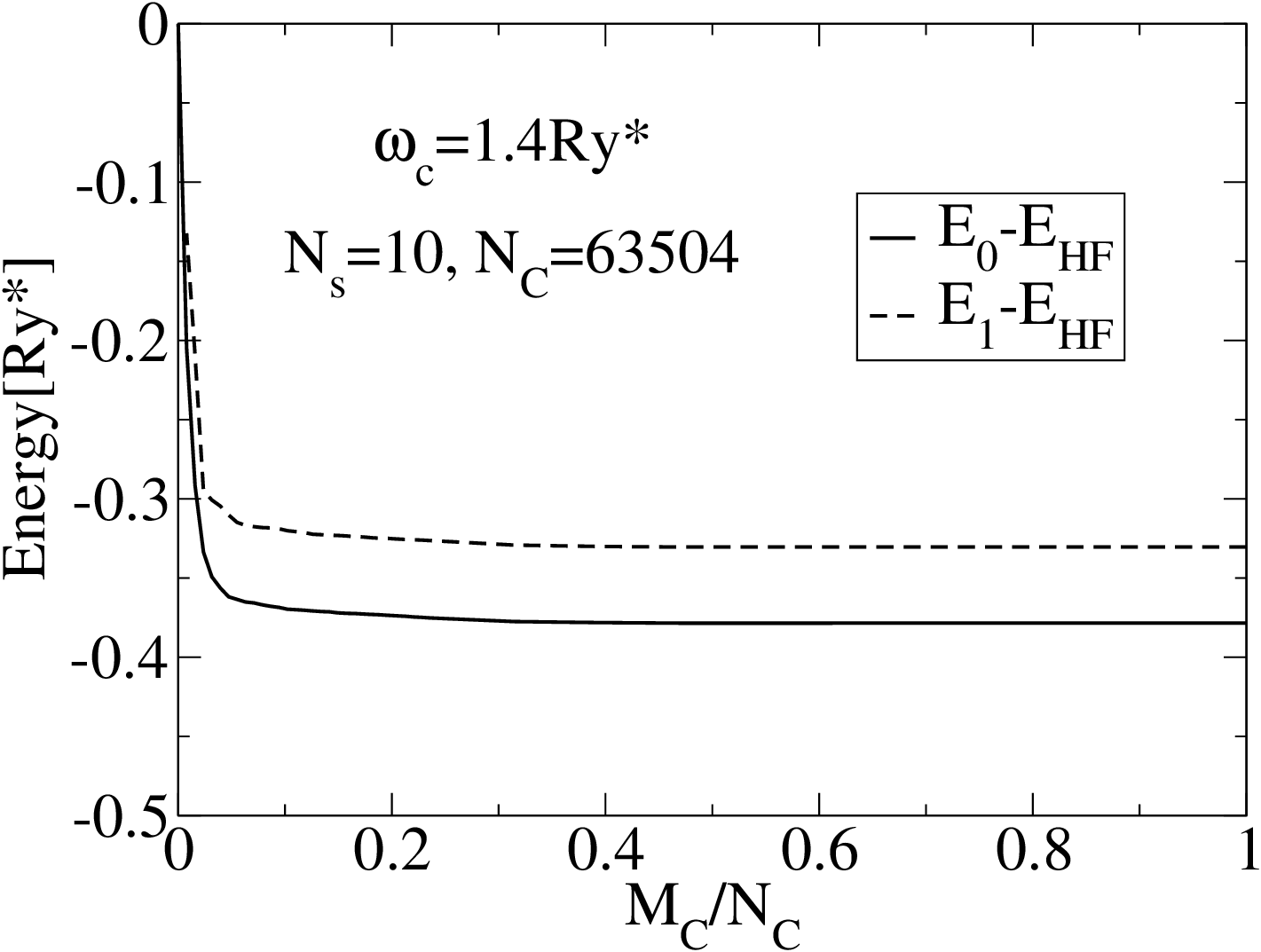}\vspace{1cm}
\includegraphics[width=0.9\linewidth]{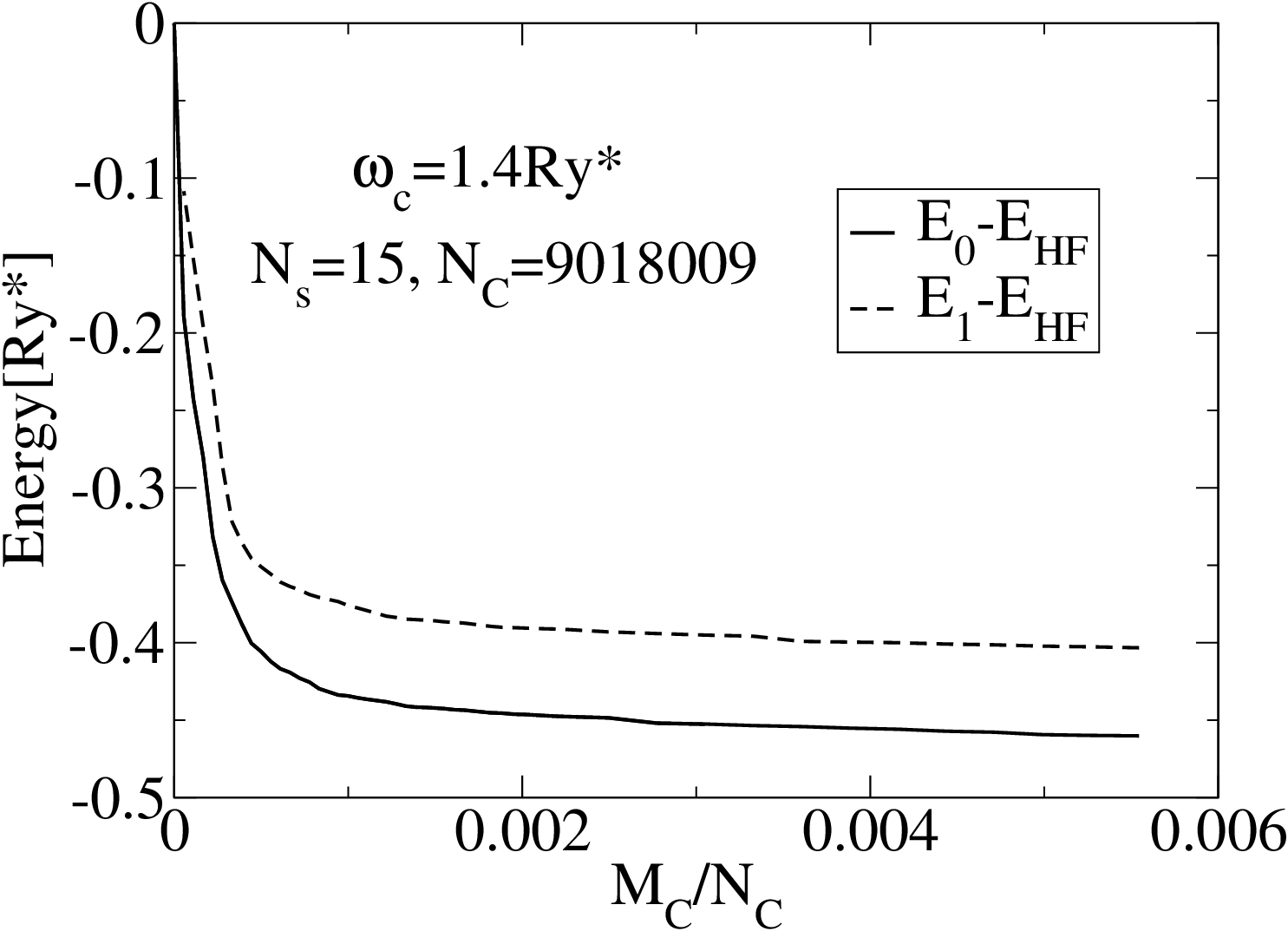}\vspace{1cm}
\noindent
\caption{
The ground state energy and the first excited state
energy in URHF-CI method is shown.
}
\label{E_conv_Just_CI15}
\end{center}
\end{figure}

\begin{figure}
\begin{center}\vspace{1cm}
\includegraphics[width=0.9\linewidth]{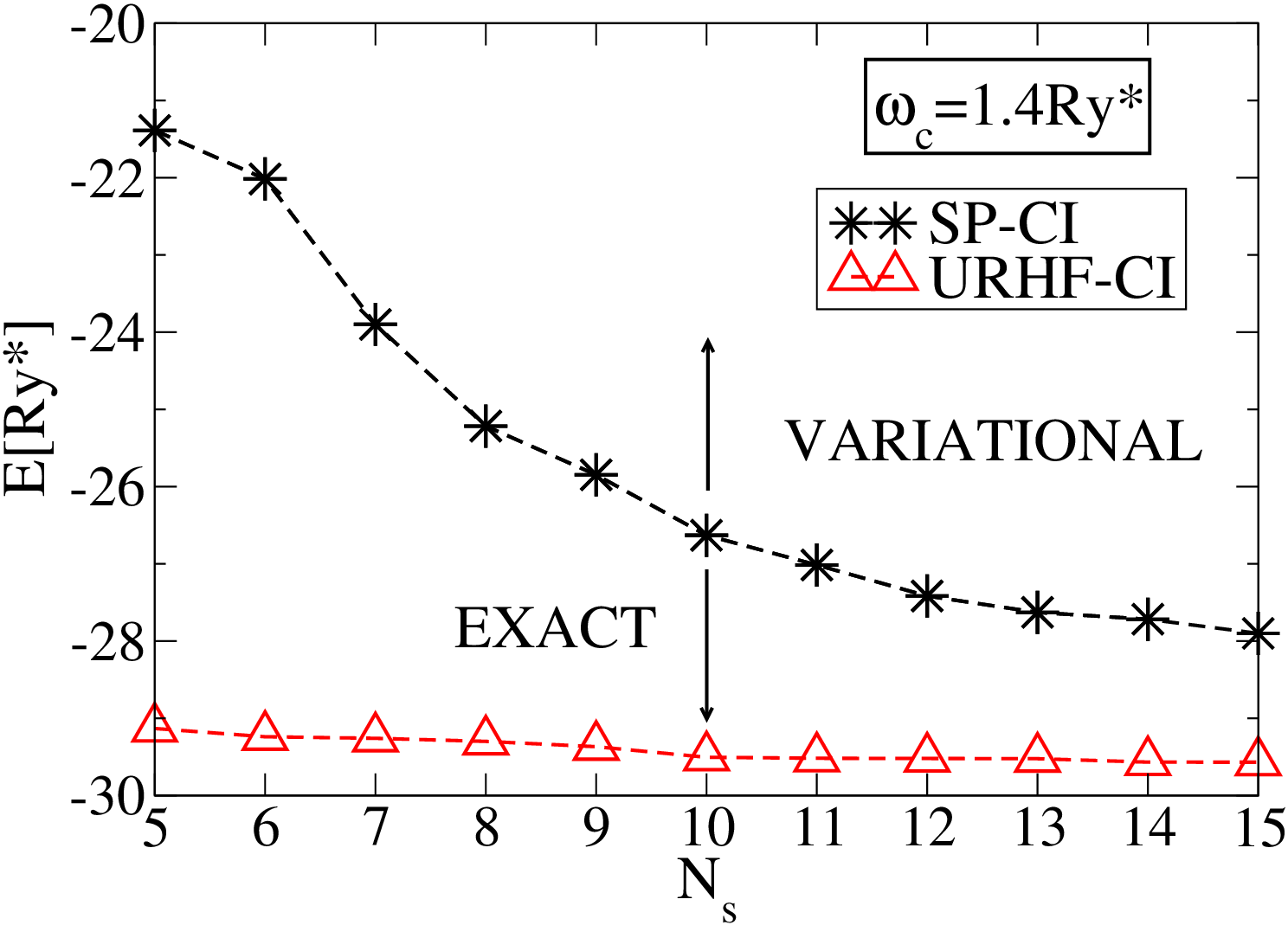}
\noindent
\caption{
The dependence of URHF-CI and SP-CI ground state energies on $N_s$.
The pre-calculated direct and exchange energies in URHF-CI improves
the ground state energy and the convergence rate.
Up to $N_s=10$ the SP-CI and URHF-CI Hamiltonian have been diagonalized
exactly.
Above $N_s=10$ the variational ground state energies are calculated
by diagonalizing $H_{\rm eff}$ with $M_C=63504$. 
}
\label{E1vsNumbConfig}
\end{center}
\end{figure}

\begin{figure}
\begin{center}\vspace{1cm}
\includegraphics[width=0.9\linewidth]{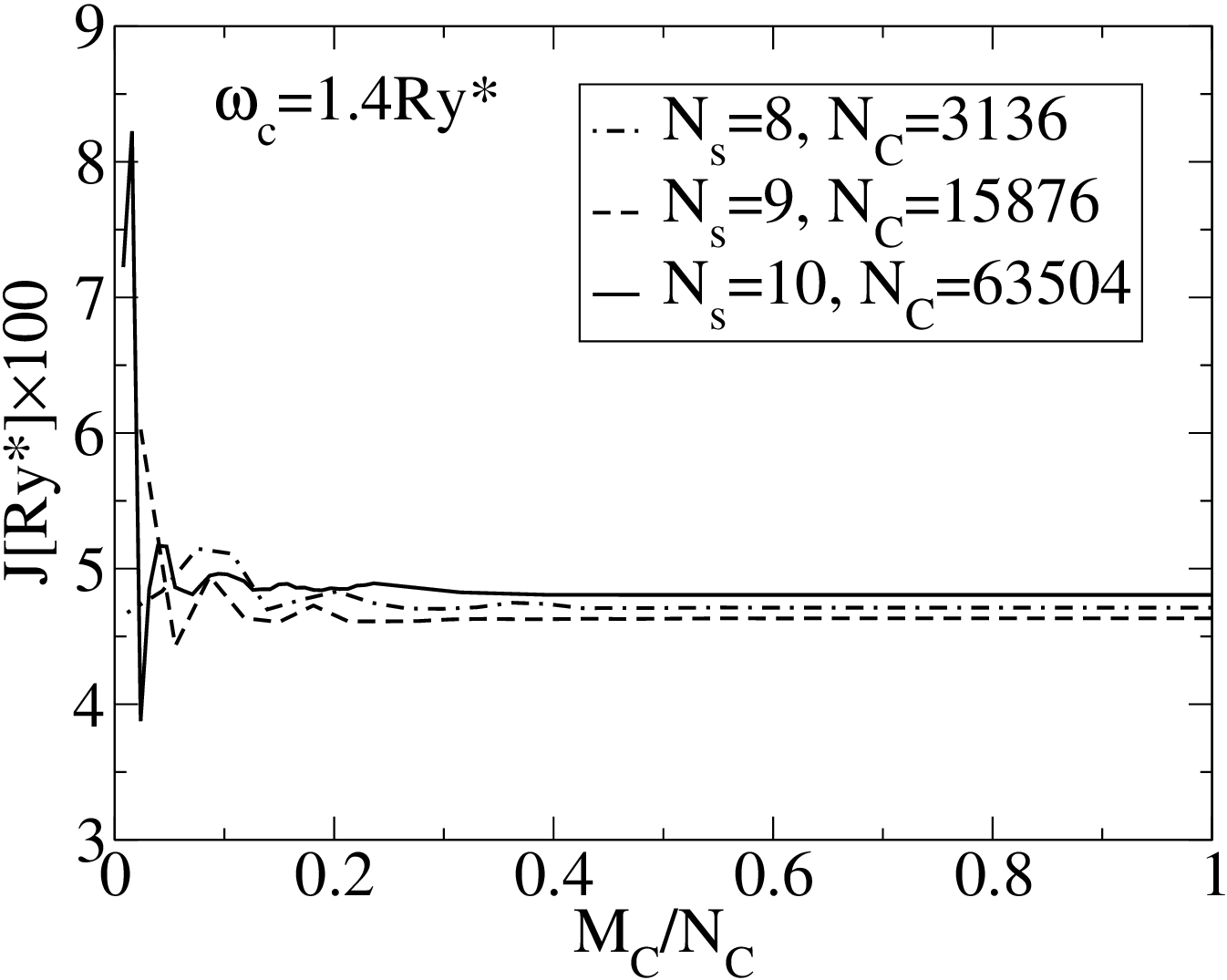}\vspace{1cm}
\includegraphics[width=0.9\linewidth]{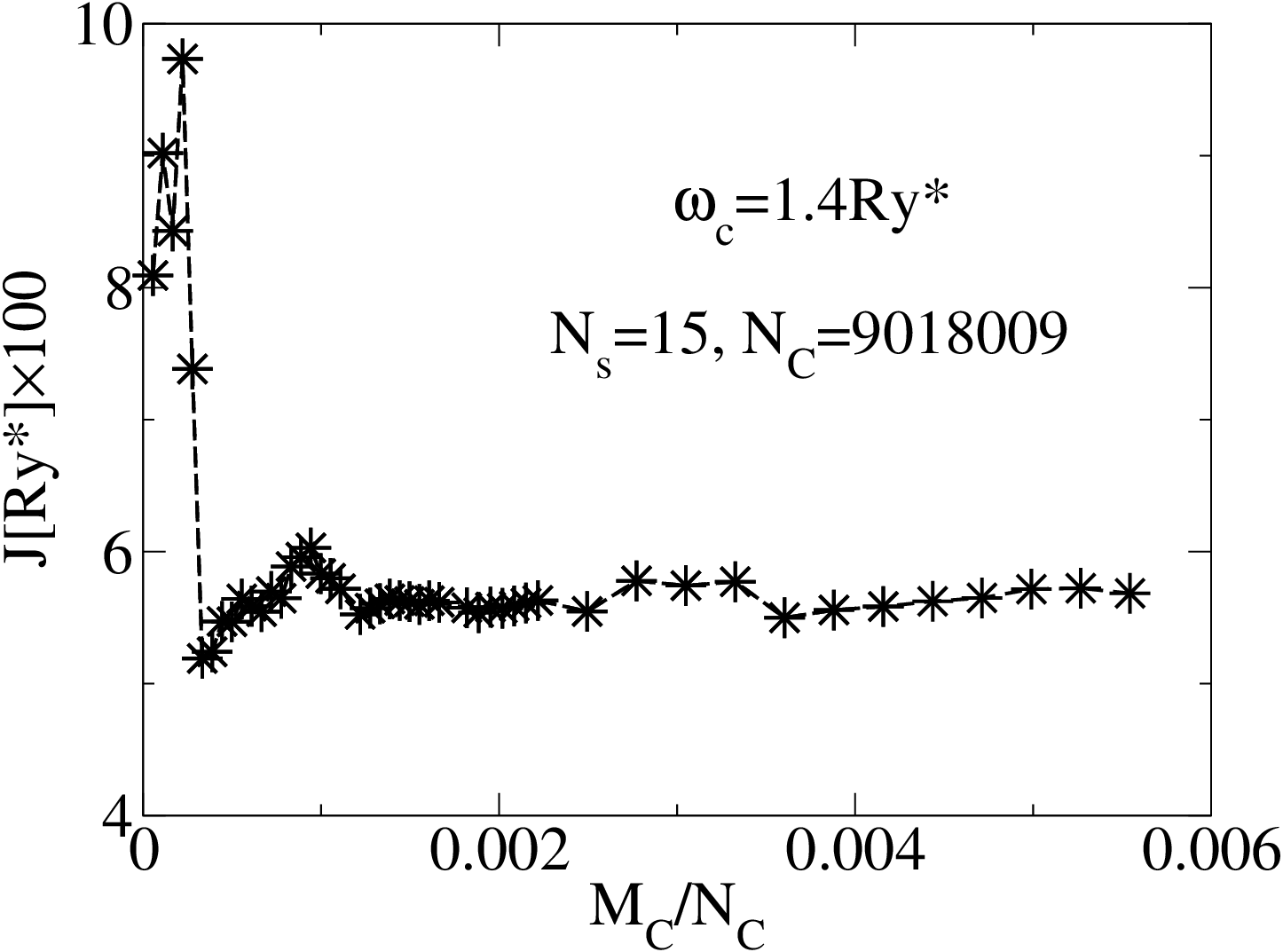}\vspace{1cm}
\noindent
\caption{
The exchange energy, the spin singlet-triplet energy gap 
in URHF-CI method is shown.
At $\hbar\omega_c=1.4Ry^*$ the spin singlet state is the ground state.
}
\label{J_conv_Just_CI15}
\end{center}
\end{figure}

\begin{figure}
\begin{center}\vspace{1cm}
\includegraphics[width=0.9\linewidth]{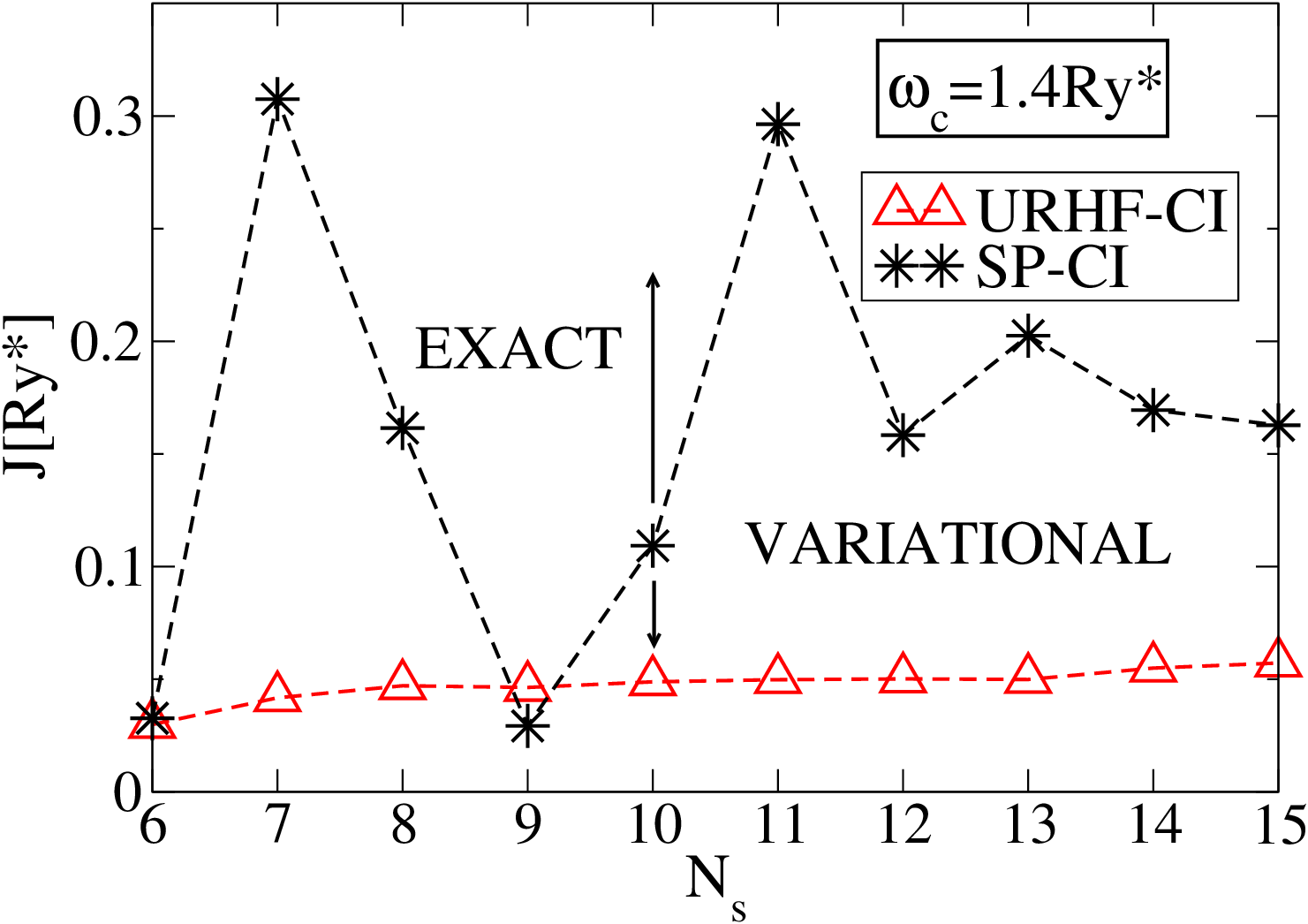}
\noindent
\caption{
A comparison between the
exchange coupling $J$ calculated in SP-CI and URHF-CI is shown.
Up to $N_s=10$ the SP-CI and URHF-CI Hamiltonian have been diagonalized
exactly.
Above $N_s=10$ a variational calculation has been used
by diagonalizing $H_{\rm eff}$ with $M_C=63504$.  
}
\label{J.vs.NumbConfig}
\end{center}
\end{figure}

In the configuration interaction method 
the Hamiltonian of an 
interacting system is calculated in the basis of finite number of
many-electron configurations. The total number of configurations 
(or Slater determinants participating in CI calculation)
is determined by
$N_C = [N_s!/N_\uparrow!(N_s-N_\uparrow)!]
[N_s!/N_\downarrow!(N_s-N_\downarrow)!]$.
Here $N_\uparrow$ and $N_\downarrow$ are the number of spin up 
and spin down electrons.
This Hamiltonian is either diagonalized exactly for small systems
or low energy eigenvalues and eigenstates
are extracted approximately for very large number of configurations.
In URHF-CI (SP-CI), $N_s$ URHF  (single particle) orbitals  are used for
constructing the Slater determinants.
By removing electrons from the occupied URHF state obtained by minimizing the
total Hartree-Fock energy and
putting them onto an unoccupied  URHF state, one can construct 
a number of configurations corresponding to electron-hole excitations.
These excitations contribute to the many body wave functions
as correlations.

Denoting the creation (annihilation) 
operators for URHF quasi-particles by $c^\dagger_{i}$ ($c_{i}$) with
the index $i$ representing the combined spin-orbit quantum numbers,
the many body Hamiltonian of the interacting system 
in the URHF basis can be written as:
\begin{equation}
H=\sum_{ij} 
\langle i | T | j \rangle 
c^\dagger_{i} c_{j} +
\frac{1}{2}\sum_{ijkl}
V_{ijkl}
c^\dagger_{i} c^\dagger_{j} 
c_{k} c_{l} ,
\label{multiparticleCI}
\end{equation}       
where 
$V_{ijkl}$ are the Coulomb matrix elements in the URHF basis.
Our method of computing Coulomb matrix elements is presented 
in the appendix. 
Here 
\begin{equation}
\langle i | T | j \rangle = \epsilon_i \delta_{ij} -
\langle i | V_H + V_X | j \rangle,
\end{equation}       
where $\epsilon_i$ are the URHF eigenenegies, and
$V_H$ and $V_X$ are the Hartree and exchange operators.
The Hamiltonian matrix is constructed in the basis of configurations
with definite $S_z$, and diagonalized using conjugated gradient methods.
As the size of URHF basis goes to infinity, the method 
becomes exact. In practice, 
one may, however, be able to obtain accurate results 
using finite number of basis in CI calculation.

To introduce a systematic method for constructing the Hamiltonian,
we select a class of configurations which
have the highest contribution to the ground state wavefunction.
Because the main contribution to the ground state energy of
many electron system comes from the Slater determinants formed
by the lowest energy HF orbitals, we accept 
Slater determinants whose HF energies are below an energy cut-off
$E_C$.
The number of such Slater determinants 
is finite, and is given by $M_C$.

The many body  Hamiltonian matrix constructed in this way is the following:
\begin{eqnarray}
H &=& \left(\begin{array}{cccccc} 
H_{11} & H_{12} &  & H_{1M_C} &  & \\
H_{21} & H_{22} &  & H_{2M_C} &  & \\
 &  &  &  &  & \\
H_{M_C1} & H_{M_C2} &  & H_{M_C M_C} &  & \\
 &  &  &  &  & \\
 &  &  &  &  & H_{N_C N_C}
\end{array}\right) \nonumber \\&&
= \left(\begin{array}{cc} 
H_{\rm eff} & G  \\
G^\dagger  & Q
\end{array}\right) \nonumber \\&&
\label{IR7}
\end{eqnarray}
Here $H_{11}<H_{22}<\dots<H_{M_C M_C}<\dots<H_{N_C N_C}$
are the HF matrix elements with $H_{11}=E_{HF}$, and $H_{M_C M_C}=E_C$
(the Coulomb interactions in the diagonal elements account
for the direct and exchange interactions).

To test this method numerically we set the cyclotron energy to $\hbar\omega_c=1.4Ry^*$,
which corresponds to the crossing and mixing of many orbitals,
 as shown in Fig. \ref{E_sp}.
Fig. \ref{DOS8} shows the evolution of URHF electron-hole excitation
spectrum as a function of cut-off $M_C$.
The energy bandwidth $W=H_{N_C N_C}-E_{HF}$ is finite.
As it is illustrated in the inset of Fig. \ref{DOS8},
the peak of URHF density of states (DOS) is in the middle of the band.
As it is shown in Fig. \ref{Scaling},
the diagonal elements of the Hamiltonian $H$ (URHF eigen-energies) 
can be represented approximately by $E_C = E_{HF} + W f(M_C/N_C)$
where $f(x)$ is a universal function.
By increasing $N_C$, the total number of URHF eigen-states $M_C$ 
(corresponding to given $E_C$) increases.
In this case, it is possible to find a lower energy variational wave function
if the energy cutoff of $H_{\rm eff}$ is set to a given $E_C$.

We now turn to investigate the convergence rate of the calculated 
ground state energy as a function of cutoff $N_C$ and $M_C$. 
In Fig. \ref{E_conv_Just_CI15} the energies, measured from the HF energy, of
the ground  and the first excited state for $N=10$ electrons obtained using
URHF basis with 
$N_s=10$ and with $N_s=15$ orbitals.
We note that for $N=10$ and $N_s=10$ the number of configurations 
$N_C= 63504$ while for $N_s=15$ and $N=10$, $N_C=9018009$. 
Hence  for $N_s=10$ the Hamiltonian can be diagonalized exactly and 
the ground and excited states are known up to $M_C=N_C$. 
This is not the case for $N_s=15$ where we were able to extract the ground 
and excited states up to $M_C=0.006 N_C$, and exact energies are not known.

However, in both cases the energies  fall off rapidly and very quickly 
saturate as a function of $M_C/N_C$.
Hence it appears  sufficient to use only a fraction of low energy 
configurations
to construct the effective Hamiltonian $H_{\rm eff}$ in order to achieve
satisfactory convergence.

Given the convergence criteria established above, we now discuss the
advantages of URHF-CI versus much easier to use SP-CI, a central result of 
this work.
Fig. \ref{E1vsNumbConfig} shows the dependence of URHF-CI and SP-CI ground 
state energy on the number
of single particle orbitals $N_s$ for $N=10$ electron droplet at 
magnetic field corresponding to $\hbar\omega_c=1.4Ry^*$.
Up to $N_s=10$ the SP-CI and URHF-CI Hamiltonian have been diagonalized
exactly.
Above $N_s=10$ the variational ground state energies are calculated
by diagonalizing $H_{\rm eff}$ with $M_C=63504$.
The lowest number of single particle orbitals populated by $N=10$ electrons 
with $S_z=0$
is $N_s=5$. For $N_s=5$ in SP-CI electrons populate the lowest five single 
particle orbitals
while in URHF electrons populate the HF orbitals, which minimize not only 
single particle energy but also direct and exchange energy. 
Hence the starting energy of initial  single configuration in URHF-CI has 
significantly lower energy compared to SP-CI. As the number of available 
states $N_s$, and hence the number of electron-hole excitations 
(configurations) increases, the ground state energy obtained in URHF-CI 
decreases very slowly. 
It starts with Hartree-Fock value of $E=-29.13 Ry^*$ for $N_s=5$
and ends up with $E=-29.6 Ry^*$ for $N_s=15$. 
This gives our best estimate of total correlation energy 
of $E_C=0.47 Ry^*$. We find the correlation energy to be only two percent 
of Coulomb energy.

By contrast with URHF-CI the  SP-CI calculations using the single 
particle basis converge very slowly. The slow convergence can be 
understood in terms of large direct and exchange energy contribution 
which the SP-CI attempts to compute very inefficiently.
Hence clear  advantage in using URHF-CI versus SP-CI method.

We now turn to the analysis of the excitation gaps. We focus on the 
energy gap between the spin singlet $S=0$ ground state and the spin triplet
$S=1$ excited state, the exchange energy $J$.

In Figs. \ref{J_conv_Just_CI15}
the convergence of calculated exchange energy $J$ 
as a function of the number of configurations $M_C/N_C$ is shown for increasing 
size $N_s$ of single particle basis. We see that increasing the size $M_C$ 
of the effective Hamiltonian initially leads to rapid oscillations in $J$  
followed by a smooth dependence. 
These calculations allow us to adjust $M_c$ to extract numerically stable 
exchange energy $J$ for each size of the single particle basis $N_s$.

This allows us to compare the dependence of calculated exchange energy
using the URHF-CI and SP-CI methods.
Fig. \ref{J.vs.NumbConfig} shows clearly fluctuations of
$J$, calculated using SP-CI method , with increasing number of
configurations. These fluctuations can be traced back to many level crossings
in single particle orbitals. We are unable to extract  reliable  value of $J$ 
using the 
commonly used SP-CI.
By contrast, exchange energy $J$ calculated using URHF-CI shows a smooth and 
convergent behavior as a function of $N_s$.
 
\subsection{HF-CI vs SP-CI - Dependence on the number of electrons N}

\begin{figure}[t]
\begin{center}\leavevmode
\includegraphics[width=0.9\linewidth]{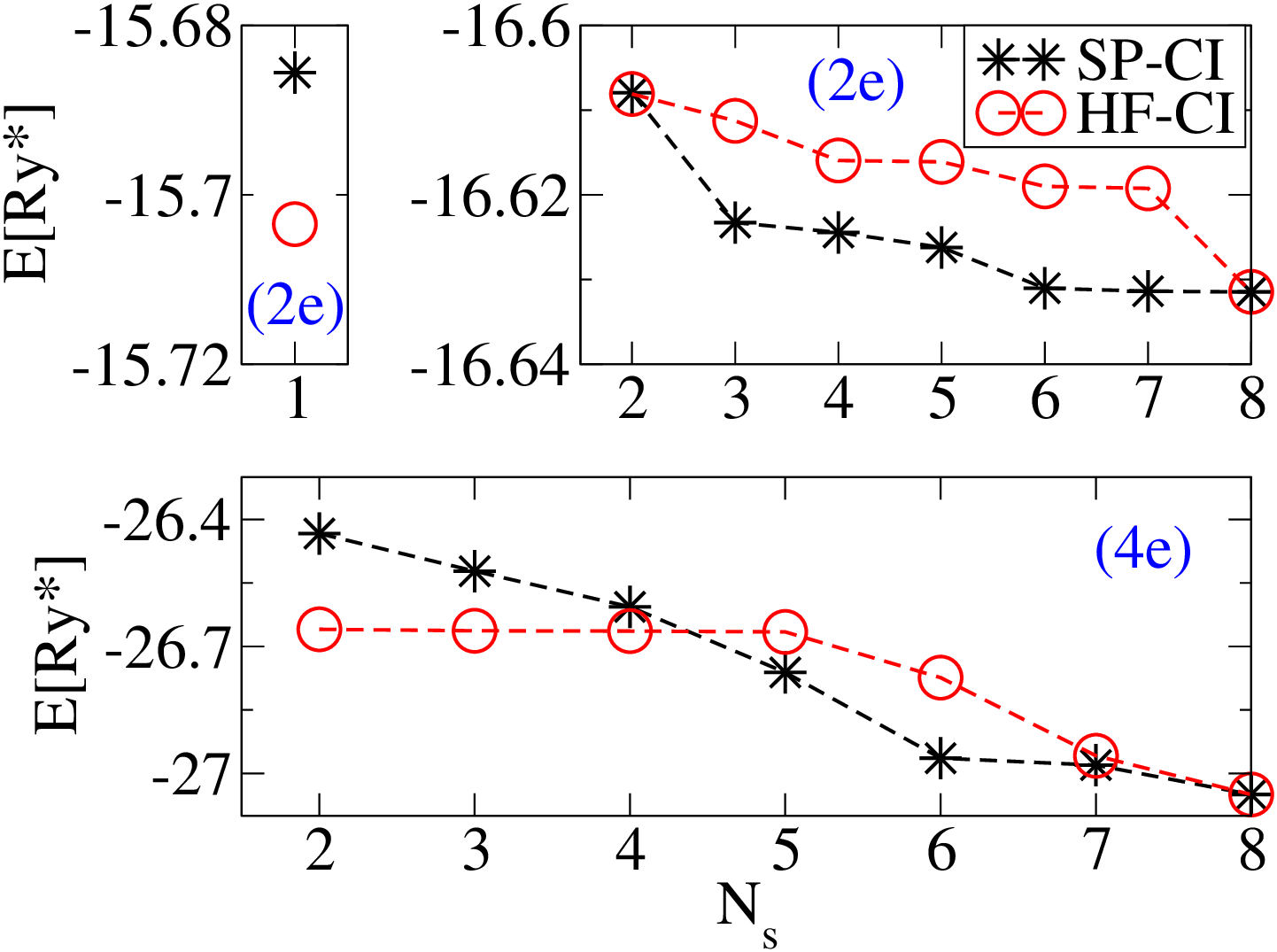}
\noindent
\caption{
Shown SP-CI and URHF-CI ground state energies 
vs. number of single particle levels used in CI calculation $(N_s)$
for two electrons (top) and 4 electrons (bottom)
at $\hbar\omega_c=1.4Ry^*$.
For two electrons, although, the ground state energy calculated from 
URHF is lower than single configuration SP-CI (top-left), but
as the number of configurations increases, SP-CI method gives
lower energy ground state (top-right).
For four electrons, the URHF-CI ground state energy is lower
than the one calculated by SP-CI for small number of configurations.
As the number of configurations increases, the ground state energy 
calculated by SP-CI becomes the lowest.
}
\label{CIvsCIHF}
\end{center}
\end{figure}

\begin{figure}[t]
\begin{center}\leavevmode
\includegraphics[width=0.9\linewidth]{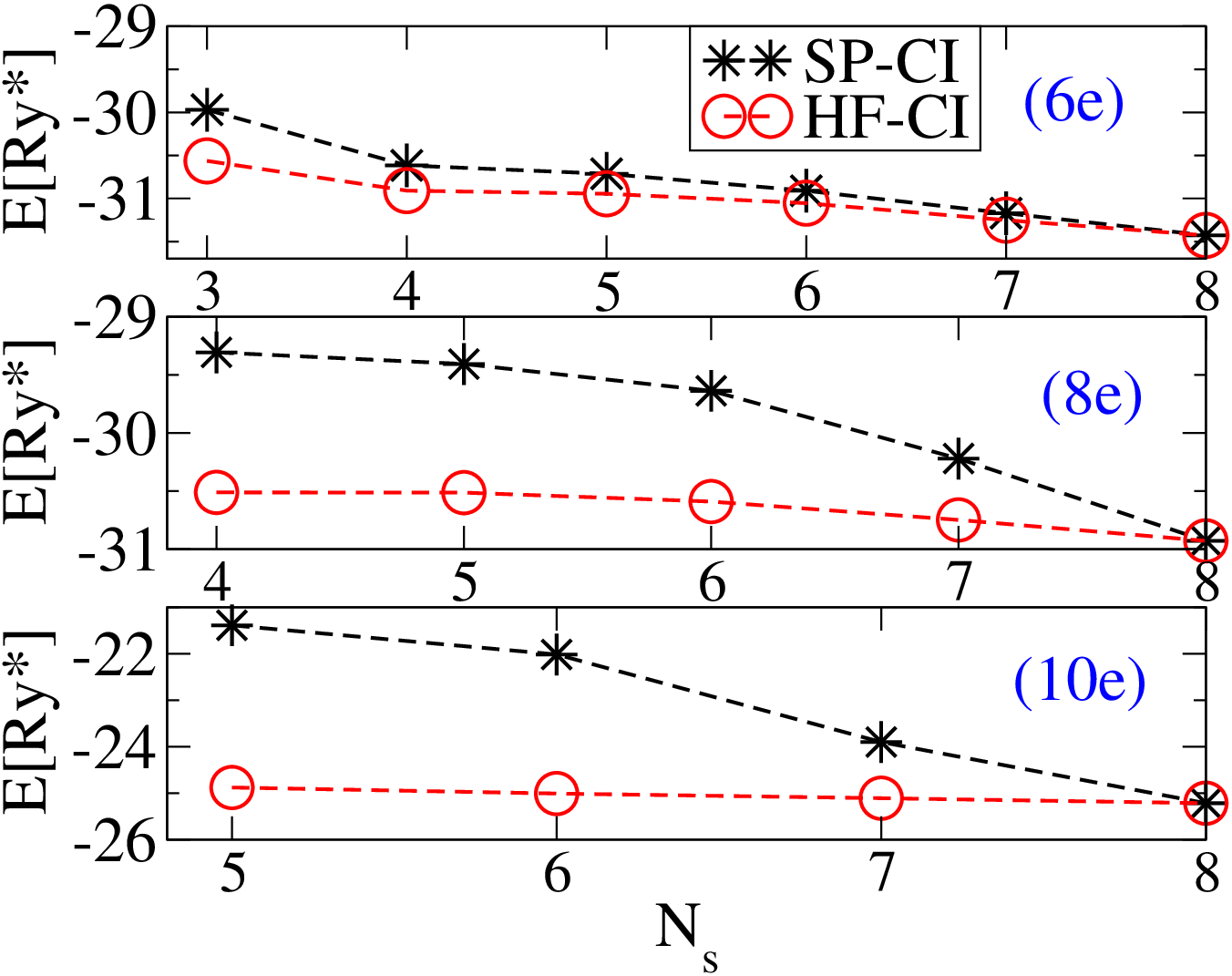}
\noindent
\caption{
A comparison between 
SP-CI and URHF-CI ground state energies 
for six (top), eight (middle) and ten electrons (bottom)
at $\hbar\omega_c=1.4Ry^*$ is shown.
In contrast to two and four electron numbers, 
URHF-CI ground state energy is lower than SP-CI ground state energy
for all configurations.
}
\label{CIvsCIHF2}
\end{center}
\end{figure}

As we discussed in last subsection, computing the spectrum of 
the CI Hamiltonian requires diagonalization of large matrices.
The size of CI Hamiltonian matrices can be optimized significantly
by a judicious choice of basis.

In preceding subsection, we introduced URHF states as a suitable basis
for CI method.
In this subsection, we present a systematic 
comparison between the ground state
energy calculated by URHF-CI and SP-CI methods to
remark that URHF-CI is a superior method to deal with a system with
large number of electrons.
We examine this in quantum dot molecules and in single dot
parabolic confining potentials.
In quantum dot molecules, we calculate the ground state energy of
a system consisting of two to ten electrons.
To compare URHF-CI and SP-CI,
one must study the behavior of the spectrum as a 
function of $N_s$, and $N_l (\geq N_s)$.
With increasing number of configurations,
this comparison is carried out up to the point that 
$N_s=N_l$, where all possible HF configurations are being exhausted.
In the limit of $N_s=N_l$, there exists 
a unitary transformation which maps 
URHF-CI Hamiltonian to SP-CI Hamiltonian, and 
thus the spectrum of SP-CI and URHF-CI become identical.
Because the number of configurations grows rapidly by increasing $N_s$,
and because we would like to reach the limit of $N_s=N_l$, 
we construct URHF states out of small number of 
single particle levels.
For the purpose of this comparison and
without any loss of generality we present the results of 
our calculation using $N_l=8$ in Figs. \ref{CIvsCIHF}-\ref{CIvsCIHF2}.
In the case of two electrons, URHF energy  
is lower than the energy of SP-CI with single Slater determinant, $N_s=1$,
(see the top-left of Fig. \ref{CIvsCIHF}).
By increasing the number of configurations we observe that
SP-CI quickly lowers the ground state energy until $N_s=8$
where URHF-CI and SP-CI become equivalent.
In the case of four electrons, the ground state energy of URHF-CI
is lower for small number of configurations.
Similar to the two electron system, with increasing number of configurations
(beyond $N_s=5$) SP-CI provides lower ground state energy.
However, with increasing number of electrons from six up to ten, 
we find that URHF-CI method gives lower ground state energy within the whole range of $N_s$, as 
 shown in
Fig. \ref{CIvsCIHF2}.

\begin{figure}[t]
\begin{center}\leavevmode
\includegraphics[width=0.9\linewidth]{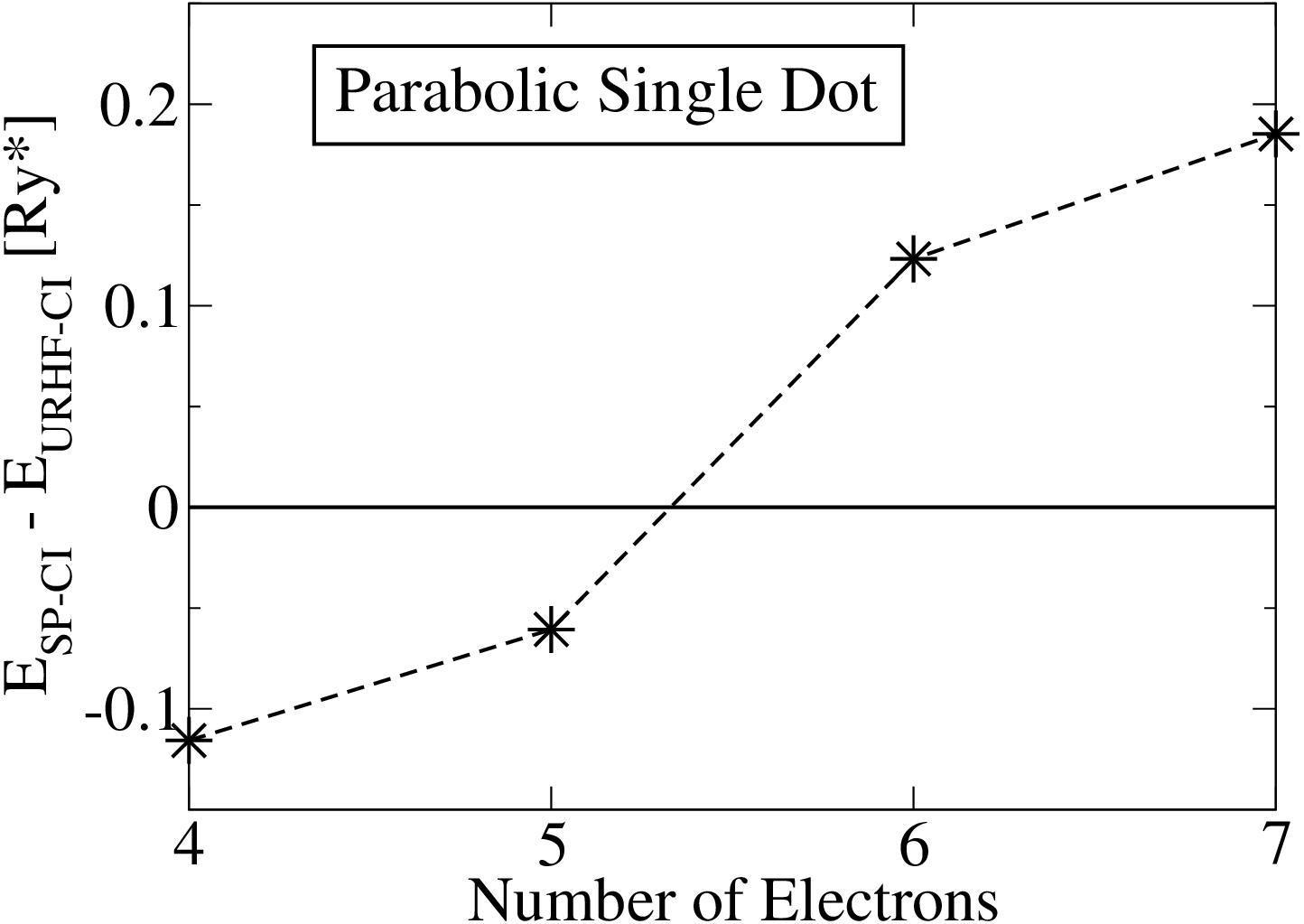}
\noindent
\caption{
Shown a zero magnetic field
comparison between GS energies calculated by SP-CI and URHF-CI methods
using $N_s=10$, $N_l=28$ for a
single dot parabolic potential ($\omega_0=0.55 Ry^*$) with
electron numbers between $N=4$, and 7.
The calculated ground state energy of 
URHF-CI (SP-CI) is lower for a system with large  (small)  
number of particles.
}
\label{CIvsCIHF_SD}
\end{center}
\end{figure}

A similar comparison can be made for electrons in a single dot with
parabolic confining potential.
The energy difference calculated by SP-CI and URHF-CI,
presented in Fig. \ref{CIvsCIHF_SD},
reveals that for electron numbers between two and five, the
ground state energy of SP-CI is lower than the ground state energy
of URHF-CI.
As expected, this order is reversed with increasing number of 
electrons ($N\geq 6$).

We therefore find  
URHF-CI method to work very well for quantum dot systems with large number of electrons.

\section{conclusion}
\label{SecVI}

We have developed real space unrestricted hybrid Hartree Fock (HF) and configuration interaction (CI) method 
(URHF-CI) suitable for the calculation
of ground and excited states of large number of electrons localized by
complex gate potentials in quasi-two-dimensional quantum dot molecules. 
 The effects of magnetic field and 
correlations are included in energies and directly in the many-particle wavefunctions
making the method an attractive candidate for potential quantum information related
applications.  
The   advantages of URHF-CI method over the commonly used CI method based on
single particle orbitals SP-CI are demonstrated.

\section{Acknowledgement}
R.A. and P.H. acknowledge the support by the NRC High 
Performance Computing project and by the Canadian Institute for Advanced Research.

\section{Appendix: Coulomb matrix elements}

In this appendix, we describe an efficient approach to calculate 
Coulomb matrix elements numerically. 
In the Hartree-Fock and configuration interaction method, 
the properties of the system are given by the single particle spectrum 
and by the Coulomb matrix elements defined as two-electron integrals 
(see Eq. [\ref{multiparticle}]). 
For the calculation involving 60 HF orbitals 
the total number of multi-dimensional integrals exceeds $12\times 10^6$.
Here we describe an efficient algorithm used in our calculation. 
In the numerical calculation of URHF-CI, the integrals of the
Coulomb matrix elements are
replaced by summation over the grids 
$\vec{r}=(ih_x,jh_y,kh_z)$  
(where $\{ijk\}$ are integers, and $\{h_x,h_y,h_z\}$
are grid spacing)
\begin{eqnarray}
V_{\alpha\beta\mu\nu} &=& \sum_{{\vec r}_1,{\vec r}_2} 
\Delta{\vec r}_1 \Delta{\vec r}_2~
\psi_{\alpha}^\ast({\vec r}_1) \psi_{\nu}({\vec r}_1) \nonumber \\&&
\frac{e^2}{\epsilon|{\vec r}_1-{\vec r}_2|}~ 
\psi_{\beta}^\ast({\vec r}_2) 
\psi_{\mu}({\vec r}_2) .
\label{Vijkl}
\end{eqnarray}  
We further transform multi-dimensional wave function 
$\psi_\alpha({\vec r})$ into a column vector $\psi_\alpha(q)$ by
mapping the  multi-dimensional vector ${\vec r}$ onto a one-dimensional 
index $\{ijk\}\rightarrow q$ where $q=1,2,\dots,N_xN_yN_z$. 
Then the multi-dimensional integral can be converted
into a vector-matrix multiplication
\begin{equation}
V_{\alpha\beta\mu\nu} = \sum_{q,q'} 
\Phi_{\alpha\nu}(q)~U(q,q')~\Phi_{\beta\mu}(q')
\label{integ_To_multi}
\end{equation}
where $\Phi_{\alpha\nu}(q) = \psi^*_\alpha(q) \psi_\nu(q)$ is a vector 
containing all the possible pairs of single-particle wave functions, and 
$U$ is the matrix with elements 
$\frac{e^2}{\epsilon|{\vec r}_1-{\vec r}_2|}$ times
$\Delta{\vec r}_1 \Delta{\vec r}_2$. 
The sums in form of do-loops can be further parallelized.
Due to the large dimension of matrix $U$, one can make use of domain 
decomposition to divide it into a number of smaller matrices 
and sum up the result of all the individual
multiplications.



\end{document}